\documentstyle[prd,aps,tighten,floats]{revtex}
\begin{document}
\draft
\title{String Equation From Field Equation}
\author{Gurovich V.Ts., 
\thanks{E-mail address: gurovich@physik.freenet.bishkek.su}
Dzhunushaliev V.D.
\thanks{E-mail address: dzhun@freenet.bishkek.su}}
\address {Theoretical Physics Department \\
          Kyrgyz State National University, Bishkek, 720024}
\maketitle
\begin{abstract}
{It is shown that the string equation can be obtain from field
equations. Such work is performed to scalar field. The equation
obtained in nonrelativistic limit describes the nonlinear string.
Such string has the effective elasticity connencted with the local 
string curvature. Some examples of the movement such nonlinear
elastic string are considered.}
\end{abstract}

The confinement phenomenon in quantum chromodynamic
is probably connected with fact that the gluon field between 
interacted quarks is
concentrated in tube stretched between this quarks. Then such picture may be 
effectively presented as a string with quarks attached on its ends. Thus,
the following problem is interesting: The string equations deduce 
from one or another field equations.
\par
In this paper we examine the simplest problem of this kind: The 
receiving of the string from nonlinear equation for the scalar field. 
As is shown in
Ref's \cite{dzh1,dzh2} the nonlinear Schr\"odinger and Heisenberg
equations have the cylindrical symmetrical solutions with the field
confined inside the tube. The field outside of such tube drops 
to zero on infinity as an exponent that gives the finite linear energy
density. This solutions make up the discrete spectrum indexed
by the number of the solution knots.
\par
In plasma theory the idea similar to the one proposed here has long been
employed. The arc discharge theory has the method reducing initial 
equations to an equation for arc discharge axis \cite{gur}.
\par
Thus, our initial equation is the nonlinear Klein-Gordon equation:
\begin{equation}
\Box \Psi = \lambda \Psi ^3 - m^2 \Psi,
\label{1}
\end{equation}
where $\Box$ is d'Alembertian; $\Psi$ is a scalar field; $\lambda$
and $m$ are some constant. Let us suppose that we have some tube solution.
We pass to the noninertial cylindrical frame of reference 
$(t,s,\rho ,\theta )$ in which the tube rests. The axis $s$ is directed
along the tube axis. For simplicity we examine the nonrelativistic
case here. Let us introduce the natural trihedron on string, then we can 
write down:
\begin{equation}
\vec r = \vec r_0 + x\vec n + y\vec b,
\label{2}
\end{equation}
where $\vec r$ is a radius vector in laboratory frame of reference;
$\vec r_0$ is a radius vector of the string (tube center); 
$\vec n$ is principal normal to a curve (to our string); $\vec b$
is a binormal ($\vec n$ and $\vec b$ are some functions on $s$);
$x,y$ are Cartesian coordinates in frame of reference connected
with natural trihedron. The polar coordinates $\rho$ and $\theta$
are introduced in the following manner:
\begin{eqnarray}
x&=&\rho \cos \theta,
\label{3}\\
y&=&\rho \sin \theta.
\label{4}
\end{eqnarray}
The line element in new coordinates looks as follows:
\begin{equation}
dl^2 = d\rho^2 + \left (1-k\rho\cos\theta\right )^2ds^2 +
\rho ^2\left (d\theta + \tau ds\right )^2,
\label{5}
\end{equation}
where $k$ is curvature of string; $\tau$ is torsion of string;
$s$ is length of string. In order to write 4-interval 
let's note that according to Einstein the presence of the week 
acceleration 
is assigned through entering in $g_{00}$ metric component the Newton 
potential. Thus, we can  write 4-interval in the following way:
\begin{equation}
dI^2 = \left (1+2g\rho\cos\theta\right )dt^2 - d\rho ^2 -
\left (1-k\rho\cos\theta\right )^2ds^2 - \rho^2\left (d\theta ^2 +
\tau ds\right )^2,
\label{6}
\end{equation}
where $g$ is acceleration in given point of the tube axis, $c=1$ is
speed of light and $\theta$ angle is measured from normal $\vec n$
in plane $\vec n, \vec b$. Further, for simlicity we consider 
the plane string ($\tau=0$).
In this frame of reference we can write down the scalar field 
near axis line in the following way:
\begin{equation}
\Psi = \Psi _0 + a\rho^2  + b\rho ^3\cos\theta + \cdots,
\label{7}
\end{equation}
where $a$ and $b$ are some functions of $s$. The substitution in field
equation (\ref{1}) gives us, with an accuracy of $\rho ^2$, the 
following equations:
\begin{eqnarray}
4a&=&m^2\Psi _0 - \lambda \Psi ^3_0,
\label{8}\\
g-k&=&0.
\label{9}
\end{eqnarray}
by $b=0$. Note that the tube solution exists only if $a<0$. Let us 
consider the left side of Eq.(\ref{9}):
$g=\partial v_n/\partial t$, where $v_n$ is normal component of speed. 
We take into account that in our frame of reference $\vec v=0$,
hence we have:
\begin{equation}
g = \frac{\partial \vec v\vec n}{\partial t} = 
\frac{\partial \vec v}{\partial t}\vec n+
\vec v \frac{\partial \vec n}{\partial t} = 
\frac{\partial \vec v}{\partial t}\vec n = a_n,
\label{10}
\end{equation}
where $a_n$ is normal component of acceleration. Finally we have the 
following equation for string (tube axis):
\begin{equation}
a_n = k.
\label{11}
\end{equation}
This equation shows us that this nonlinear string has some effective
elasticity connected with local curvature of the center line of the tube. 
Later we examine several simple examples of the movement of the
such string. It is easy shown that this equation coincides with
standart equation for string \cite{wit} by small speed and bending:
\begin{equation}
\Box x^{\mu} = 0.
\label{11a}
\end{equation}
where $x_{\mu}$ are spasetime coordinates of the strings.
At the present time there are some arguments that the string has some 
effective elasticity connected with its finite thickness 
\cite{pol}, \cite{kl}, \cite{ger}.
\par
Let us consider the closed string rotating around its center 
and which is a circle. We came to frame of reference rotating together
with the string. The form string is defined by the function $\rho (t)$,
where $\rho$ is the string radius. Then Eg.(\ref{11}) looks as  
follows:
\begin{equation}
\ddot\rho = -\frac{1}{\rho} + 
\frac{\left (L/2\pi\lambda\right )^2}{\rho ^5},
\label{12}
\end{equation}
where $L$ is a string angular momentum and $\lambda$ is linear
mass density of the string; the second term in right side is
centrifugal acceleration. Here is also taken into account \cite{wit} 
that the linear string density is invariable by extension, contraction
and deformation in according with generally accepted string
model. In this case the energy conservation law can be written down:
\begin{equation}
\frac{1}{2}\dot\rho^2 + \ln\frac{\rho}{\rho _0} +
\frac{\left (L/2\pi\lambda\right )^2}{4\rho ^4} = E.
\label{13}
\end{equation}
The form of potential energy from Eq.(\ref{13}) indicates that
the string executes an anharmonic oscillation about equilibrium
state $\rho _0 = (L/2\pi\lambda )^{1/2}$.
\par
Let us consider the rotation of the open string around the line
passing through its ends. We introduce the Cartesian coordinate 
system with $x$ axis directed along the rotation axis. Thus, 
the string form is defined as function $y(x,t)$. Hence, 
Eq.(\ref{11}) has the following view:
\begin{equation}
\frac{y_{tt}}{\left (1+y^2_x\right )^{1/2}} =
\frac{y_{xx}}{\left (1+y^2_x\right )^{3/2}} +
\frac{\left (L/J\right )^2y}{\left (1+y^2_x\right )^{1/2}},
\label{14}
\end{equation}
where $\omega = L/J$ is rotation frequency of the string ($L$ is
string angular momentum, $J=\lambda\int y^2ds$ is moment of 
inertia in given time moment) and taking into account that  
the centrifugal acceleration onto normal is to be projected. 
We will seek the equilibrium configuration. In this case we 
derive the equation for $y(x)$:
\begin{equation}
\frac{y_{xx}}{1+y^2_x} = -\omega ^2y.
\label{15}
\end{equation}
Which has the following solution:
\begin{equation}
x = \int\limits_y^{y_0}\left (a_0\exp \left \{-\omega ^2_0y^2\right \}
-1 \right )^{-1/2}dy.
\label{16}
\end{equation}
\par
Let us consider the finite string the ends of which move 
along $y$ axis with some acceleration. In this case Eq.(\ref{11})
can be written down in the following way:
\begin{equation}
y_{tt} = \frac{y_{xx}}{1+y^2_x},
\label{17}
\end{equation}
where Cartesian $x$ coordinate parametrizes the string point; the string 
movement takes place along $y$ axis. We seek the solution in the
following form:
\begin{equation}
y(x,t) = a(t) + b(x),
\label{18}
\end{equation}
This makes possible to separate the variables and we receive the 
following solution:
\begin{equation}
y(x,t) = g\frac{t^2}{2} + v_0t + y_0 - 
\frac{1}{g}\ln\cos\left (2gx\right ).
\label{19}
\end{equation}
where $y_0$ is a initial point and $v_0$ is a initial speed of string.
The string sags at small acceleration $g<\pi/2l$, $l$ is the string 
length. There is a critical acceleration ($g=\pi/2l$) 
in which string is torn (its length is infinity).
\par 
More detailed examination of given model of the nonlinear elastic
string will be presented in the extended version of this paper.

\end{document}